\documentclass[a4paper]{article}
\usepackage{INTERSPEECH2022}
\usepackage[dvipsnames]{xcolor}

\usepackage{graphicx}
\usepackage{float}
\usepackage{cuted}
\usepackage{acronym}

\acrodef{SOTA}{State-Of-The-Art}
\acrodef{SI-SNR}{Scale-Invariant Signal-to-Noise Ratio}
\acrodef{SNR}{Signal-to-Noise Ratio}
\acrodef{GLU}{Gated Linear Unit}

\usepackage{algorithm}
\usepackage{algpseudocode}

\title{Deep Audio Waveform Prior}
\name{Arnon Turetzky, Tzvi Michelson, Yossi Adi, Shmuel Peleg}

\address{
School of Computer Science and Engineering\\
  The Hebrew University of Jerusalem}
\email{\{arnon.turetzky, tzvi.michelson\}@mail.huji.ac.il}
\begin{document}

\maketitle

\begin{abstract}
Convolutional neural networks contain strong priors for generating natural looking images \cite{ulyanov2018deep}. These priors enable image denoising, super resolution, and inpainting in an unsupervised manner. Previous attempts to demonstrate similar ideas in audio, namely deep audio priors, (i) use hand picked architectures such as harmonic convolutions, (ii) only work with spectrogram input, and (iii) have been used mostly for eliminating Gaussian noise \cite{zhang2019deep}. In this work we show that existing \ac{SOTA} architectures for audio source separation contain deep priors even when working with the raw waveform. Deep priors can be discovered by training a neural network to generate a single corrupted signal when given white noise as input. A network with relevant deep priors is likely to generate a cleaner version of the signal before converging on the corrupted signal.  We demonstrate this restoration effect with several corruptions: background noise, reverberations, and a gap in the signal (audio inpainting).
\end{abstract}
\vspace{0.25cm}

\section{Introduction}
It has been demonstrated that the success of deep convolutional neural networks in computer vision is due to deep priors in the convolutional architecture itself and should not be attributed to the training process alone \cite{ulyanov2018deep}. Leveraging these priors, it is possible to perform image denoising, super resolution, inpainting, and more in an unsupervised fashion without any pretraining \cite{ulyanov2018deep, gandelsman2019double}. These discoveries helped bridge the gap between classical methods and machine or deep learning methods, while providing insights into the inner workings of deep architectures.

Zhang et al. \cite{zhang2019deep} showed that classical architectures do not provide good priors for audio. They examined the classical WaveUnet, and showed that even working with spectrograms, regular and dilated convolutions suffer from severe limitations. For example, when trying to fit three noised stationary signals of 1,000, 2,000 and 3,000 Hz the networks fitted the signal and the noise at an equal pace. To overcome these limitations, they proposed the usage of harmonic convolutions. Their goal was to exploit harmonic structures as an inductive bias for auditory signal modeling. Despite the success of their experiments, current SOTA audio architectures do not utilize harmonic structures and tend to rely more on standard building blocks.

Later, Tian et. al. \cite{tian2019deep, tian2020deep} demonstrated that in certain cases deep spectrogram priors can still be found in regular and dilated convolutions. Using varying dilation schedules and interwoven skip connections, Narayanaswamy et al. succeeded in strengthening these priors \cite{narayanaswamy2021design}.
Michelashvili et al. \cite{michelashvili2019speech} showed that it is possible to preform speech denoising in an unsupervised manner without priors by leveraging the inherent lack of structure in noisy signals. In their work, they identified the pixels in the spectrogram which fluctuate most during the training of the network. Using the intensity of the fluctuations they created a mask over the spectrogram which can be used to differentiate between the original signal and the noise. 

Recently, D{\'e}fossez et al. \cite{defossez2020real} proposed the Demucs model for the task of music source separation. The Demucs model is a neural network composed of a fully convolutional encoder followed by a sequential modeling over the encoder's output, and a reverse decoder with U-net like connections between the encoder and decoder (we describe the model later in more detail). This model reached SOTA performance in music source separation \cite{defossez2019music, defossez2021hybrid} and speech enhancement \cite{defossez2020real}. In this work we show that using this new architecture it is possible to preform denoising, dereverberation, and inpainting on the raw waveform in an unsupervised manner using priors inherent to the network.

\begin{figure}[t!]
\centering 
\includegraphics[width=\columnwidth]{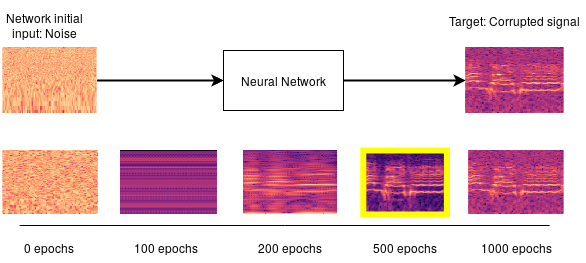}
\caption{
Visualization of the deep prior process for denoising. A Neural Network is trained to generate a target corrupted signal from an initial input which is pure noise. At 500 epochs the generated signal is substantially cleaner then at 1000 epochs, when the corrupted signal is recreated. Network was applied on raw waveform, and spectrograms are used only for visualization.}
\label{fig:visual_deep_prior}
\vspace{-16pt}
\end{figure}

\begin{figure}[t]
  \centering
  \vspace{-22pt}
  \includegraphics[width=\linewidth]{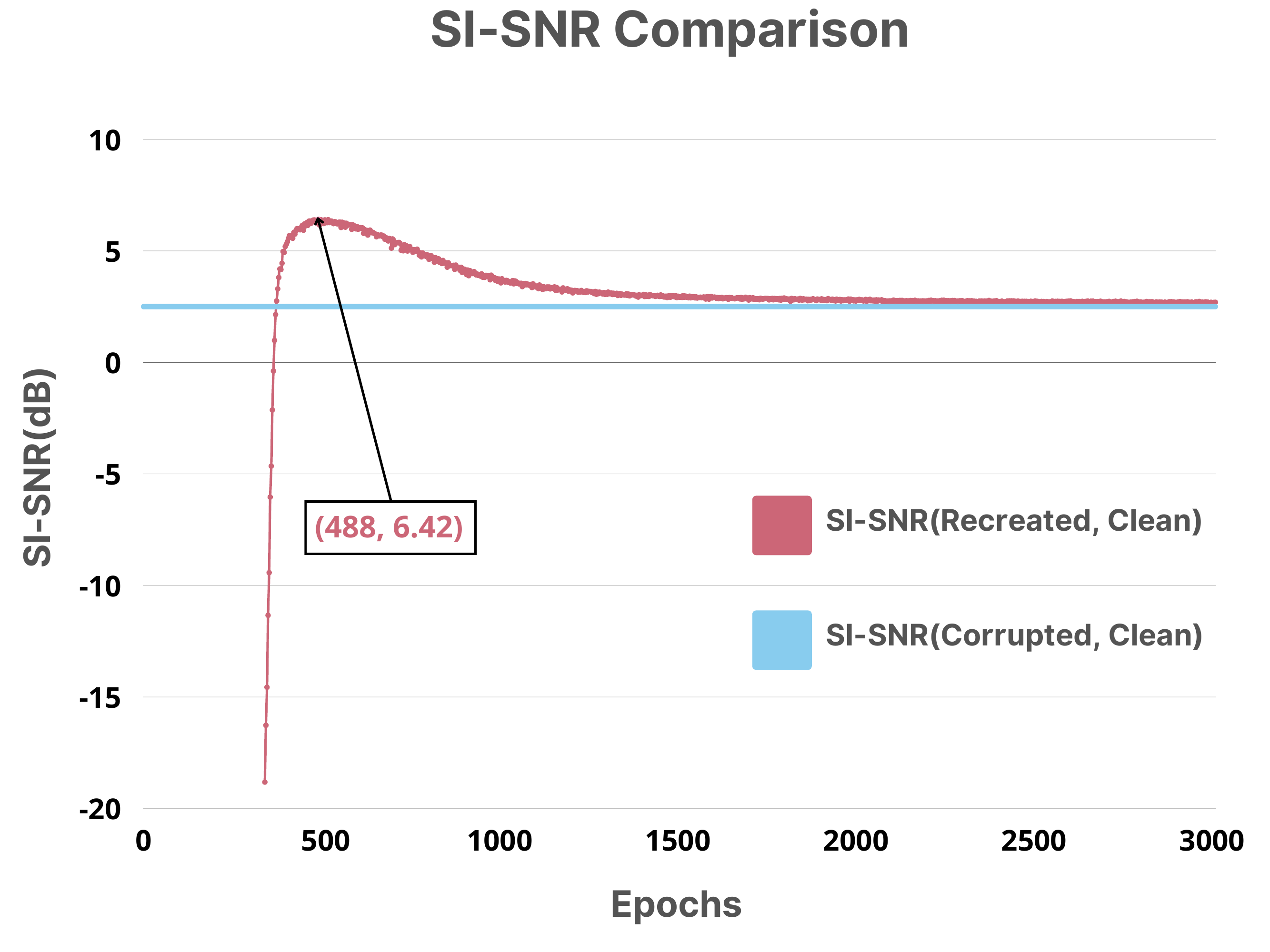}
  \caption{Graph of \ac{SI-SNR} as a function of epochs. We can see that the SI-SNR(Recreated, clean) goes above the \ac{SI-SNR}(Corrupted, Clean) meaning the recreated signal becomes more similar to the clean signal then the corrupted signal.}
  \label{fig:SI-SNR}
  \vspace{-16pt}
\end{figure}

\section{Deep Waveform Priors Overview}
To discern whether a network contains deep priors relevant to a specific input, we fit a generator network to a single corrupted signal. The network weights are randomly initialized, and fitted to the signal using gradient descent. Thus, the network weights serve as a parametrization for the signal. In this manner, the only information used to perform reconstruction is contained in the single corrupted input signal and the architecture of the generator network. If the architecture of the network contains relevant deep priors, it is possible that during training, the generator will give a cleaner signal before overfitting the corrupted version. Figure~\ref{fig:visual_deep_prior}
presents a diagram showing this process. In the diagram we see that after five hundred epochs the signal is cleaner then the original noisy signal. The graph in Figure~\ref{fig:SI-SNR} shows the behavior of the \ac{SI-SNR} during the process described above. In the graph, the line representing SI-SNR(recreated,  clean) goes unexpectedly above the line representing SI-SNR(corrupted, clean) before leveling downwards. This essentially means that the recreated signal becomes more similar to  the clean signal then the corrupted one.

We use the waveform Demucs architecture \cite{defossez2019music} as the basis for our analysis. The Demucs architecture consists of a series of down-sampling 1D convolution layers each followed by a \ac{GLU}, a number of LSTM layers, and then a series of up-sampling 1D deconvolution layers each followed by a \ac{GLU}. Following Ulyanov et al. \cite{ulyanov2018deep} we removed the skip connections from the network. Intuitively, this makes space for the network's inductive priors to work by propagating the loss only from the very end of the network. When the skip connections remain in place, they correct the weights to overfit the noisy target in every part of the network, and do not leave the network the freedom necessary for the priors to work. In order to allow the network to learn despite the removal of the skip connections it was necessary to reduce the number of convolutional and LSTM layers.

\subsection{Implementation details}
In different experiments different variants of the architecture were used, to achieve improved results, though a systemic evaluation of all the variants was done only in the ablation studies. The Adam optimizer was used throughout all the experiments with a learning rate of 1e-4. The networks were trained using L1 loss to calculate the difference from the raw waveforms. The input noise was sampled from a Gaussian distribution with an std of 0.1, and was resampled every epoch in all experiments excluding the audio inpainting, where it was only sampled once. We used a kernel of size 8 with a step of 4 similarly to the default Demucs architecture. In all experiments the networks were trained until convergence (see Figure~\ref{fig:SI-SNR}), though almost all converged within 3000 epochs. The audio was sampled at 16Khz in all sections, for all audio types to allow for fair comparison.

\begin{figure}[t]
  \vspace{-20pt}
  \centering    
  \includegraphics[width=\linewidth]{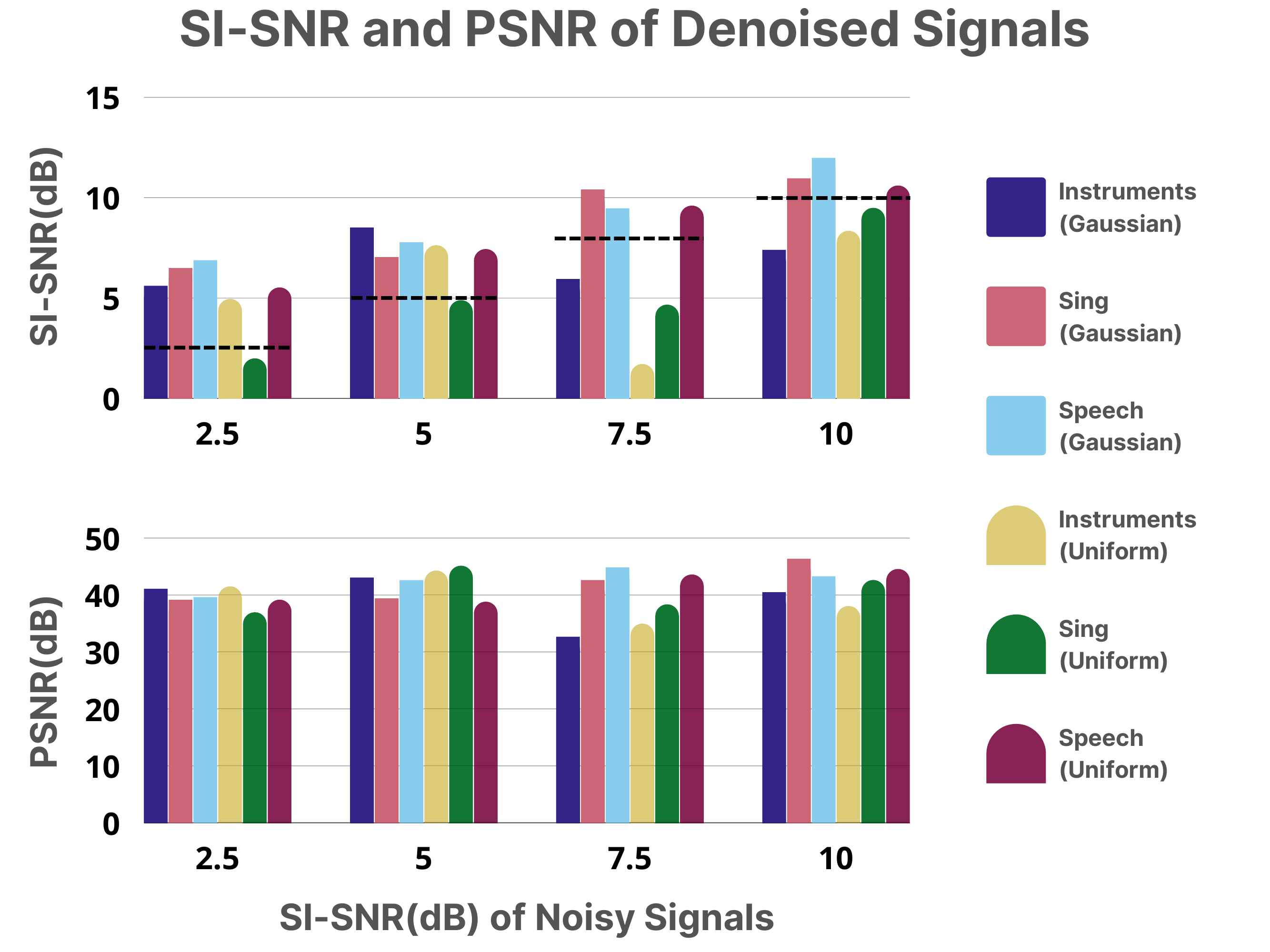}
  \caption{This figure shows the SI-SNR and PSNR of the denoised singals. Each bar represents the average of the metric maximum (similarly to the one marked in Figure~\ref{fig:SI-SNR}) over 20 randomly sampled tracks of appropriate class, noise type and noise level. We can see the limits of the deep priors as the SI-SNR increases and it is harder to differentiate between the signal and the noise.}
  \label{fig:denoising}
\end{figure}

\subsection{Metrics}
We use \ac{SI-SNR}, PSNR and PESQ as evaluation functions. \ac{SI-SNR} is defined as in \cite{nachmani2020voice, chazan2021single} using the following formula:
\begin{equation}
 \text{SI-SNR}(s,\hat{s}) = 10\ log_{10}\ \frac{\|\tilde{s} \|^2}{\| e \|^2},
\end{equation}
where \(\ \tilde{s}=\frac{\langle s,\hat{s} \rangle s }{\|s\|^2}\), \( e=\hat{s}-\tilde{s} \). We consider PSNR as follows: 
\begin{equation}
\text{PSNR}(s,\hat{s})=10\ log_{10}\left(\dfrac{\text{MAX}_{I}^2}{\text{MSE}}\right),
\end{equation}
where $MAX_I$ is the difference between the maximum and the minimum amplitude values. To calculate PESQ we use the python-pesq package\footnote{https://github.com/ludlows/python-pesq.}.



\section{Denoising}
\begin{figure}[t]
  \centering
  \includegraphics[width=\linewidth]{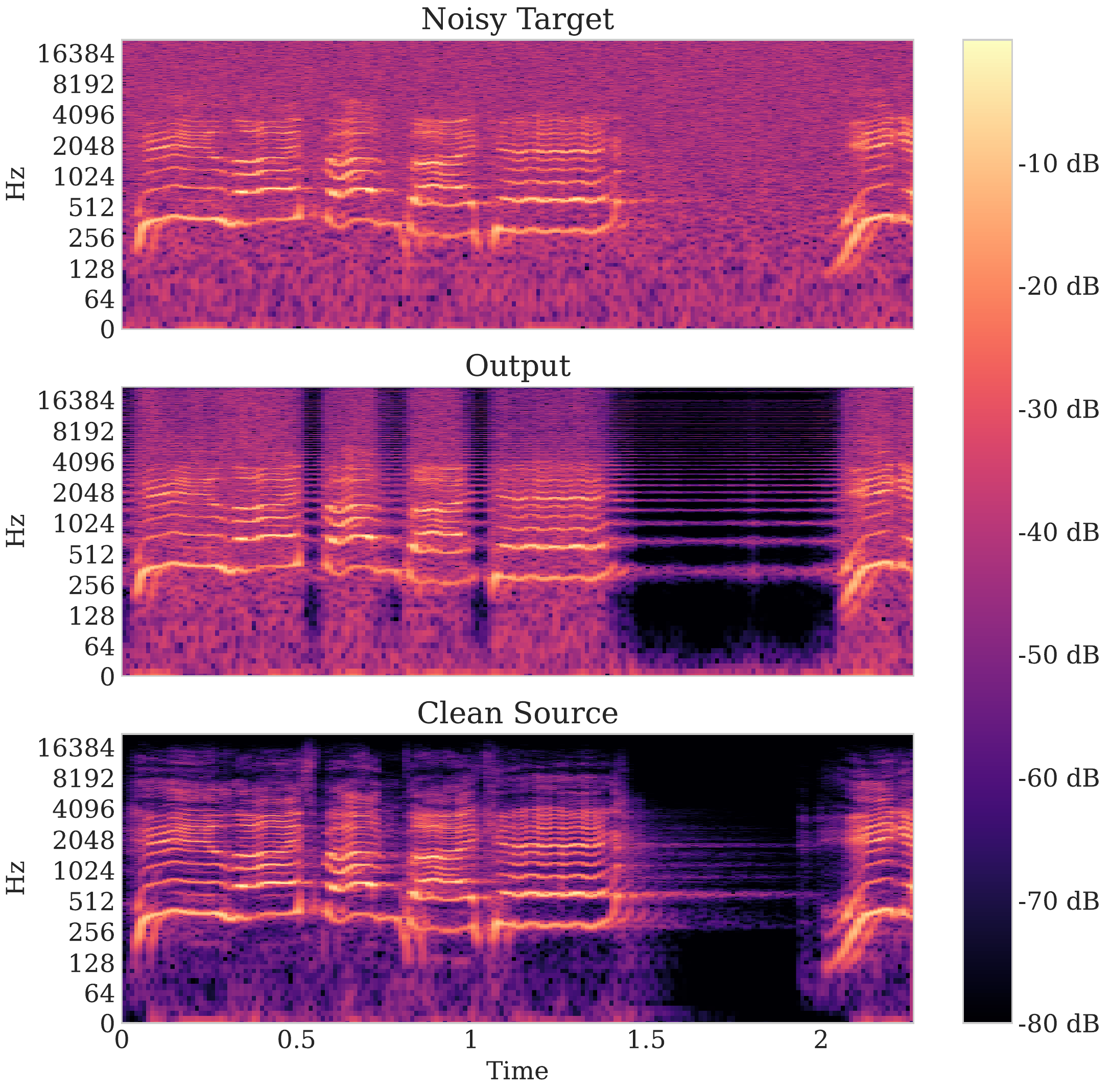}
  \caption{The spectrograms of (a) Noisy target; (b) Network output; (c) Original clean source. As the network recreates the raw waveform, spectrograms are used only for visualization.}
  \label{fig:Spectrograms}
\end{figure}

The first type of corruption we attempt to correct is denoising. We corrupt the signal using two types of noise:
\begin{itemize}
    \item \textbf{Gaussian noise} --- sampled from a Gaussian distribution and added to every sample in the raw waveform
    \item \textbf{Uniform noise} --- sampled from a Uniform distribution and added to every sample in the raw waveform
\end{itemize}
The noise is added to a variety of sounds: 
\begin{itemize}
    \item \textbf{Speech} --- taken from Valentini et al.'s dataset \cite{valentini2017noisy}
    \item \textbf{Singing} --- taken from the MUSDB18 dataset \cite{musdb18}
    \item \textbf{Instruments} --- taken from MedleyDB 2.0 \cite{medley}
\end{itemize}
Throughout this paper all our clean audio is taken from these 3 datasets. To allow for comparisons between different noise types the noises are added at equal intensities. The addition of the noise was done using code published by Xia et al. \cite{9054254}. Thus, we create noisy samples with a number of different \ac{SI-SNR}s and allow comparison between them. 
The results of our analysis can be found in Figure~\ref{fig:denoising}. The graphs report the average maximum metrics achieved by the output signal while training the neural network to generate the corrupted signal. While these results are not competitive with SOTA denoising methods, they do show the existence of a deep prior inherent to the architecture which is the goal of this study. For every level of noising, that is SI-SNR(corrupted, clean), we can see that the network usually performs some denoising before fitting the noisy signal.  Figure~\ref{fig:Spectrograms} visualizes one of our denoising results using spectrograms. 

\section{Dereverberation}
Reverberations are acoustical noise appearing in enclosed spaces through multiple reflections of the sound on the walls and objects of a room. When a speaker talks in a room, these multiple echoes add to the direct sound and blur its temporal and spectral characteristics \cite{lebart2001new}. Reverberant speech can be described as sounding distant with noticeable echos. These detrimental perceptual effects generally increase with increasing distance between the source and the microphone \cite{habets2007single}. In this work we attempt to clean the reverberations from the signal treating them as we treated other types of noise in the previous section. This task is inherently more difficult then the other denoising tasks we performed since the corruption itself has the same structure as the signal to be cleaned. As expected, our success here was more limited although a modest amount of dereverberation was obtained. A summary of our experimental results can be found in Table~\ref{tab:sound_dereverberation_results}. We used the pyroomacoustics python package to add reverberations to our audio \cite{2018} as well as Valentini et al.'s dataset \cite{valentini2017noisy}. Every setting in the table was averaged over 20 samples, although the results were robust even when averaging over only 10-15 samples.

To better evaluate the level of dereverberation improvement, results reported in Table ~\ref{tab:sound_dereverberation_results} are the SI-SNR improvement metric (SI-SNRi) over the reverberant signal.
We can see that the priors succeed in removing reverberations to a certain extent. PESQ is not reported in the table as it was very unstable and with most samples achieved very high scores which did not represent an improvement in audio quality. Our hypothesis to explain the results in the table is as follows. When the reverberation is very weak (Rt60=0.1) the audio sounds very similar to the original, so there is very little to improve, as such the improvements achieved are minor. When the reverberations are of medium strength (Rt60=0.2) there are more reverberations to clean, yet they can still be distinguished from the original signal, so the network has a stronger effect. When Rt60=0.5 the reverberations are too strong for the prior to differentiate between them and the original signal, and the network recreates them both equally well.

\begin{table}[tb]
\centering
\small
\caption{Dereverberating results for different input signals with different Rt60s. Rt60 is defined as the number of seconds it takes for the reverberations to decrease by 60 dB. The numbers reported in the table are the average \ac{SI-SNR} and PSNR improvement over the mixture that the network's output managed to achieve relative to the reverberated signal.}
\begin{tabular}{llcc}
\toprule
\bf RT60(sec) & & \bf SI-SNRi & \bf PSNRi \\
\midrule
0.1 & Speech & 2.22 & 13.45\\
& Singing & 1.15 & 14.40\\
& Instruments & 1.15 & 14.14\\
\midrule
0.2 & Speech & 4.76 & 16.30\\
& Singing & 5.48 & 18.45\\
& Instruments & 5.79 & 18.45\\
\midrule
0.5 & Speech & -0.11 & 5.66\\
& Singing & -0.08 & 6.65\\
& Instruments & -0.03  & 6.45\\
\bottomrule
\end{tabular}
\vspace{0.25cm}
\label{tab:sound_dereverberation_results}
\end{table}

\section{Audio Inpainting}
\begin{figure}[t!]
  \centering
  \includegraphics[width=3.1in]{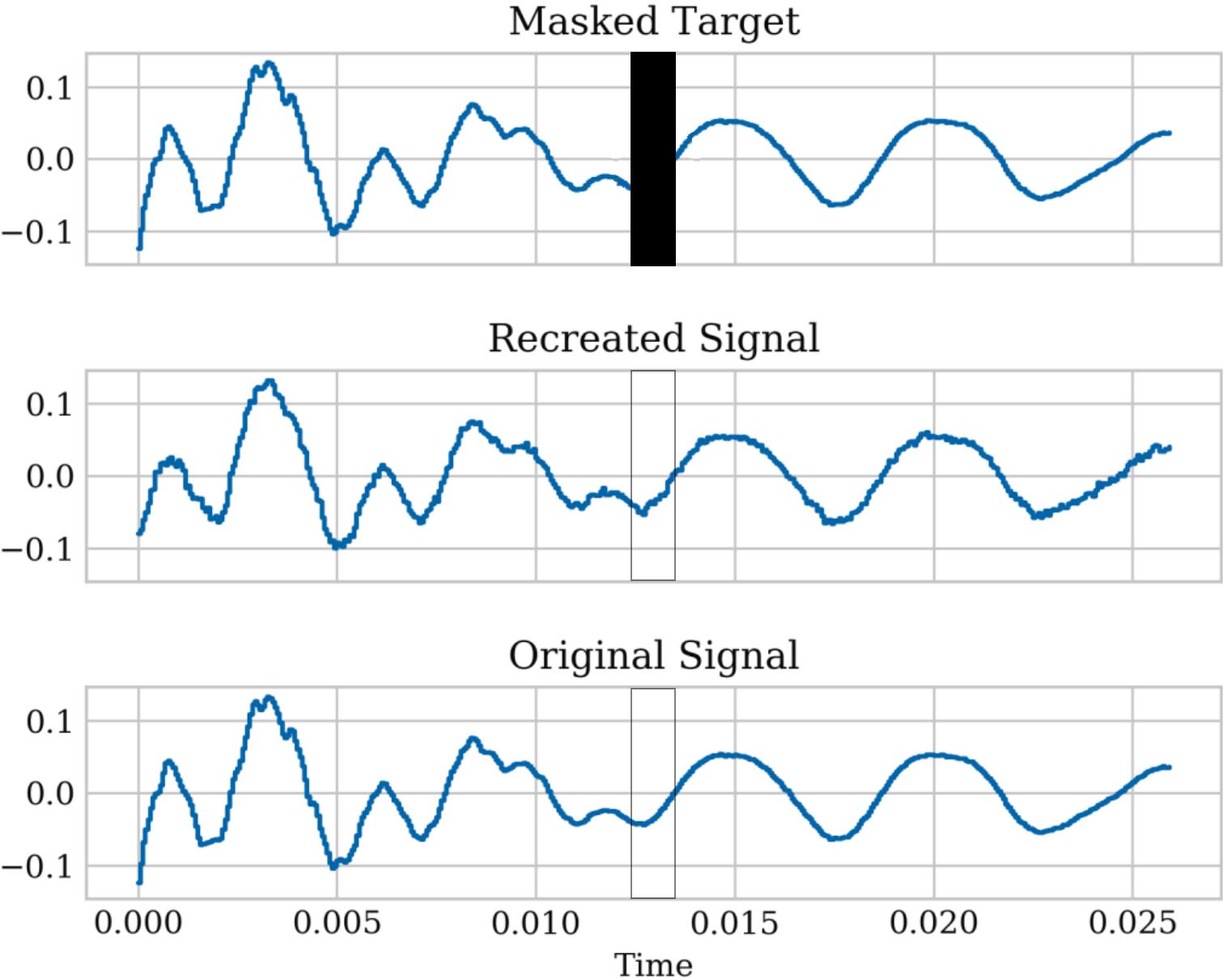}
  \caption{Waveforms of masked signal, recreated signal and original (un-masked) signal. The masked area (the horizontal line in the circle) is recreated and the waveform of the recreated signal is very close to the waveform of the original signal.}
  \label{fig:inpainting}
\end{figure}

Inpainting is a classical problem in computer vision and has been used to demonstrate deep priors of convolutional networks. When performing inpainting a mask is placed over part of the image and the network recreates the image behind the mask. In audio, speech and music signals are often subject to localized distortions, where the intervals of distorted samples are surrounded by undistorted samples. Examples may include noises or clicks, CD scratches, old recordings, packet loss in cellphones etc. \cite{adler:inria-00577079}. Audio inpainting in the waveform domain contains two additional challenges:
\begin{itemize}
    \item \textbf{Dimensionality} -- at a standard sample rate of 16KHz masking even 10 ms affects 160 samples. Hence, the network must recreate these 160 consecutive samples.
    \item \textbf{2D vs. 1D} -- When recreating an image context can be taken from 360 degrees, as the picture is 2-dimensional. When recreating a waveform, context can only be taken from a single dimension.
\end{itemize}
In this work we masked a small (1-5 ms, see Table~\ref{tab:audio_inpainting_results}), non-silent, segment of a 2 second clip, and trained the network to recreate the input signal. The training loss ignored the network's output under the mask. Hence, the network is unobstructed in recreating this part of the signal and it's deep priors can come into view. A visualization of our results can be found in Fig.~\ref{fig:inpainting}. The figure shows that the network recreates a good approximation of the original signal, despite no loss being calculated on the masked area. 

To measure the impact of the network priors we compare the recreation quality of the Demucs network to the recreation quality of WaveUnet which has been shown to contain very week priors \cite{michelashvili2019speech, zhang2019deep}. A summary of the results can be found in Table~\ref{tab:audio_inpainting_results}. The results reported are averaged over 20 clips each one of 2 seconds. The masked segment is randomly sampled within the clip. The table shows a number of points. First, the Demucs architecture consistently succeeds in recreating the signal  better then the WaveUnet. Second, as the length of the masked signal increases the quality of the recreation goes down. Third, the Demucs architecture is less successful at recreating samples of singing. It is interesting to note that the Demucs architecture achieved SOTA results in music separation and speech enhancement, but has not been used for singing audio, similar to our results.

\begin{table}[tb]
\centering
\small
\caption{The table describes the metrics reached by the network when inpainting the masked signal using the Demucs architecture vs. using the WaveUnet architecture (which contains substantially weaker priors). Every result is averaged over 20 clips with the masked section randomly sampled within the clip. We can see that the Demucs architecture is consistently superior to WaveUnet in all metrics. However, as the length of the mask grows the quality of the inpainting decreases.}
\begin{tabular}{llcccc}
\toprule
 & & \multicolumn{2}{c}{\textbf{SI-SNR}} & \multicolumn{2}{c}{\textbf{PSNR}}  \\
\midrule
& & WUnet & Demucs & WUnet & Demucs \\
\midrule
1ms & Speech & -2.05 & \textbf{6.57} & 12.02 & \textbf{32.71}\\
& Music & -1.83 & \textbf{6.81} & 11.97 & \textbf{34.46}\\
& Singing & -3.73 & \textbf{4.21} & 11.95 & \textbf{38.40} \\
\midrule
2ms & Speech & -5.33 & \textbf{3.55} & 10.53 & \textbf{34.02}\\
& Music & -5.77 & \textbf{2.98} & 10.79 & \textbf{30.44}\\
& Singing & -8.45 & \textbf{-2.35} & 10.02 & \textbf{31.64} \\
\midrule
5ms & Speech & -12.13 & \textbf{-2.60} & 9.73 & \textbf{30.46}\\
& Music & -9.48 & \textbf{-2.11} & 10.47 & \textbf{30.03}\\
& Singing & -11.07 & \textbf{-4.60} & 10.15 & \textbf{31.46} \\
\bottomrule
\end{tabular}

\vspace{0.25cm}
\label{tab:audio_inpainting_results}
\end{table}

\begin{table}[t!]
\centering
\small
\caption{Ablation studies done on different variations of the Demucs architecture. When no activation function is mentioned a standard ReLU is used.}
\begin{tabular}{lccc}
\toprule
\textbf{Architecture} & \textbf{SI-SNR} & \textbf{PSNR} & \textbf{PESQ} \\
\midrule
Conv (2 layers) & 3.05 & 39.12 & 1.38\\
Conv (4 layers)& 3.07 & 37.29 & 1.76\\
Conv (6 layers)& 2.64 & 38.36 & 1.33\\
Conv + skip & 2.97 & 36.07 & 1.61 \\
Conv + GLU & 2.85 & 36.54 & 1.42\\
\textbf{Conv + LSTM} & \textbf{5.42} & \textbf{40.56} & \textbf{2.31}\\
Conv + Attention & 3.26 & 40.37 & 1.99\\
Conv + GLU + LSTM & 5.01 & 39.64 & 1.85\\
Conv + GLU + Attention & 2.94 & 37.38 & 1.96\\

\bottomrule
\end{tabular}
\vspace{0.25cm}
\label{tab:ablation}
\end{table}

\section{Ablation Studies}
To better understand which parts of the Demucs architecture cause the architecture to have deep priors we analyzed the network from the bottom up. At each step we added another element to the network and saw how this affects the prior. Our analysis was done using randomly sampled Uniform noise with an SI-SNR of 2.5dB to allow the differences between the variants of the architecture to be seen clearly. With each variant of the architecture we randomly sampled 20 audio clips, denoised them, and averaged the results. Table~\ref{tab:ablation} reports our results. When the activation is not reported a standard ReLU is used. The original architecture is the Conv + LSTM + GLU option, we added attention layers instead of LSTM layers as part of our ablation studies to see their possible effect. Additionally, we examined the effects of different amounts of convolutional layers as well as the effect of skip connections. Using an LSTM alone did not provide results so the option is not reported in the table. There are a number of points we can learn from the table:
\begin{itemize}
    \item GLU improves the priors of the network when used with convolutional layers.
    \item When used with the LSTM layers the effect of the GLU is detrimental.
    \item Adding LSTM layers to the convolutional layers improves the network's prior.
    \item Adding attention layers instead of LSTM layers does not improve the network's priors.
\end{itemize}
We understand these results to mean that convolutional layers, aside from containing deep priors themselves, are necessary to prepare the waveform for the sequential modeling (LSTM) and to revert the sequential model back to the waveform. However, the LSTM itself contains significant deep priors. Additionally, although this is not represented in the table, Conv + LSTM works better with only 2 layers of convolution, and not with four. We hypothesize, that as the network deepens the learning becomes harder due to vanishing gradients. Skip connections can not be used as a solution, since they do not leave the network sufficient freedom to learn.

\section{Conclusions}
In this work we show for the first time (to the best of our knowledge)
the power of deep waveform priors in a \ac{SOTA} audio architecture. We demonstrate the strength of these priors on 3 separate tasks: audio denoising, audio dereverberation, and audio inpainting and achieve results which could not be achieved without deep priors. We believe, the findings presented in this study shed light on the recent success of the Demucs architecture \cite{defossez2019music} on several source separation tasks operating over the raw-waveform.


\vspace{6pt}
\noindent \textbf{Acknowledgements:} This research was supported by grants from the Israel Science Foundation, the DFG, and the Crown Family Foundation.

\bibliographystyle{IEEEtran}

\begin{thebibliography}{10}
\providecommand{\url}[1]{#1}
\csname url@samestyle\endcsname
\providecommand{\newblock}{\relax}
\providecommand{\bibinfo}[2]{#2}
\providecommand{\BIBentrySTDinterwordspacing}{\spaceskip=0pt\relax}
\providecommand{\BIBentryALTinterwordstretchfactor}{4}
\providecommand{\BIBentryALTinterwordspacing}{\spaceskip=\fontdimen2\font plus
\BIBentryALTinterwordstretchfactor\fontdimen3\font minus
  \fontdimen4\font\relax}
\providecommand{\BIBforeignlanguage}[2]{{%
\expandafter\ifx\csname l@#1\endcsname\relax
\typeout{** WARNING: IEEEtran.bst: No hyphenation pattern has been}%
\typeout{** loaded for the language `#1'. Using the pattern for}%
\typeout{** the default language instead.}%
\else
\language=\csname l@#1\endcsname
\fi
#2}}
\providecommand{\BIBdecl}{\relax}
\BIBdecl

\bibitem{ulyanov2018deep}
D.~Ulyanov, A.~Vedaldi, and V.~Lempitsky, ``Deep image prior,'' in \emph{CVPR},
  2018, pp. 9446--9454.

\bibitem{zhang2019deep}
Z.~Zhang, Y.~Wang, C.~Gan, J.~Wu, J.~B. Tenenbaum, A.~Torralba, and W.~T.
  Freeman, ``Deep audio priors emerge from harmonic convolutional networks,''
  in \emph{Int. Conf. Learning Representations (ICLR)}, 2019.

\bibitem{gandelsman2019double}
Y.~Gandelsman, A.~Shocher, and M.~Irani, ``"double-dip": Unsupervised image
  decomposition via coupled deep-image-priors,'' in \emph{CVPR}, 2019, pp.
  11\,026--11\,035.

\bibitem{tian2019deep}
Y.~Tian, C.~Xu, and D.~Li, ``Deep audio prior,'' \emph{arXiv:1912.10292}, 2019.

\bibitem{tian2020deep}
------, ``Deep audio prior: Learning sound source separation from a single
  audio mixture,'' \emph{CVPR Workshops}, 2020.

\bibitem{narayanaswamy2021design}
V.~S. Narayanaswamy, J.~J. Thiagarajan, and A.~Spanias, ``On the design of deep
  priors for unsupervised audio restoration,'' \emph{arXiv:2104.07161}, 2021.

\bibitem{michelashvili2019speech}
M.~Michelashvili and L.~Wolf, ``Speech denoising by accumulating per-frequency
  modeling fluctuations,'' \emph{arXiv:1904.07612}, 2019.

\bibitem{defossez2020real}
A.~D{\'e}fossez, G.~Synnaeve, and Y.~Adi, ``Real time speech enhancement in the
  waveform domain,'' in \emph{Interspeech}, 2020.

\bibitem{defossez2019music}
A.~D{\'e}fossez, N.~Usunier, L.~Bottou, and F.~Bach, ``Music source separation
  in the waveform domain,'' \emph{arXiv:1911.13254}, 2019.

\bibitem{defossez2021hybrid}
A.~D{\'e}fossez, ``Hybrid spectrogram and waveform source separation,''
  \emph{arXiv:2111.03600}, 2021.

\bibitem{nachmani2020voice}
E.~Nachmani, Y.~Adi, and L.~Wolf, ``Voice separation with an unknown number of
  multiple speakers,'' in \emph{Int. Conf. Machine Learning (ICML)}, 2020, pp.
  7164--7175.

\bibitem{chazan2021single}
S.~E. Chazan, L.~Wolf, E.~Nachmani, and Y.~Adi, ``Single channel voice
  separation for unknown number of speakers under reverberant and noisy
  settings,'' in \emph{ICASSP 2021-2021 IEEE International Conference on
  Acoustics, Speech and Signal Processing (ICASSP)}.\hskip 1em plus 0.5em minus
  0.4em\relax IEEE, 2021, pp. 3730--3734.

\bibitem{valentini2017noisy}
C.~Valentini-Botinhao, ``Noisy speech database for training speech enhancement
  algorithms and tts models,'' University of Edinburgh. School of Informatics.
  Centre for Speech Technology Research, Tech. Rep., 2017.

\bibitem{musdb18}
\BIBentryALTinterwordspacing
Z.~Rafii, A.~Liutkus, F.-R. St{\"o}ter, S.~I. Mimilakis, and R.~Bittner, ``The
  {MUSDB18} corpus for music separation,'' Dec. 2017. [Online]. Available:
  \url{https://doi.org/10.5281/zenodo.1117372}
\BIBentrySTDinterwordspacing

\bibitem{medley}
R.~Bittner, J.~Wilkins, H.~Yip, and J.~Bello, ``Medleydb 2.0: New data and a
  system for sustainable data collection,'' in \emph{Int. Conf. on Music
  Information Retrieval (ISMIR-16)}, 2016.

\bibitem{9054254}
Y.~{Xia}, S.~{Braun}, C.~K.~A. {Reddy}, H.~{Dubey}, R.~{Cutler}, and
  I.~{Tashev}, ``Weighted speech distortion losses for neural-network-based
  real-time speech enhancement,'' in \emph{ICASSP 2020 - 2020 IEEE
  International Conference on Acoustics, Speech and Signal Processing
  (ICASSP)}, 2020, pp. 871--875.

\bibitem{lebart2001new}
K.~Lebart, J.-M. Boucher, and P.~N. Denbigh, ``A new method based on spectral
  subtraction for speech dereverberation,'' \emph{Acta Acustica united with
  Acustica}, vol.~87, no.~3, pp. 359--366, 2001.

\bibitem{habets2007single}
E.~Habets, ``Single- and multi-microphone speech dereverberation using spectral
  enhancement,'' Ph.D. dissertation, Technische Universiteit Eindhoven, 2007.

\bibitem{2018}
R.~Scheibler, E.~Bezzam, and I.~Dokmanic, ``Pyroomacoustics: A python package
  for audio room simulation and array processing algorithms,'' \emph{ICASSP},
  Apr 2018.

\bibitem{adler:inria-00577079}
A.~Adler, V.~Emiya, M.~G. Jafari, M.~Elad, R.~Gribonval, and M.~D. Plumbley,
  ``Audio inpainting,'' \emph{IEEE Trans. on Audio, Speech and Language
  Processing}, vol.~20, no.~3, pp. 922 -- 932, Mar. 2012.

\end{thebibliography}


\end{document}